\documentclass[a4paper]{article}
\usepackage{graphicx}
\usepackage{fullpage}
\usepackage{latexsym}
\usepackage{mathrsfs}
\usepackage{amssymb,amsbsy}
\usepackage{amsfonts}
\usepackage{caption}
\usepackage{subcaption}
\usepackage{float}
\usepackage{wrapfig}
\usepackage{epstopdf}
\usepackage{multirow}

\newcommand{\as}{\prime\prime}
\renewcommand{\deg}{^{\circ}}

\date{}
\begin{document}

\title{Measurements of Gondola Motion on a Stratospheric Balloon Flight}

\author{Margarita Safonova, K.~Nirmal, A.~G. Sreejith, Mayuresh Satpodar, \\
Ambily Suresh, Ajin Prakash,
Joice Mathew, Jayant Murthy, \\
(Indian Institute of Astrophysics, Bangalore), \\
 Devarajan Anand, B.~V.~N.~Kapardhi, B.~Suneel Kumar, P.~M.~Kulkarni,\\
(Tata Institute of Fundamental Research, Balloon Facility, Hyderabad)}

\maketitle

\begin{abstract} 

Balloon experiments are an economically feasible method of conducting observations in astronomy that are not possible from the ground. The astronomical payload may include a telescope, a detector, and a pointing/stabilization system.  Determining the attitude of the payload is of primary importance in such applications, to accurately point the detector/telescope to the desired direction. This is especially important in generally unstable lightweight balloon flights. However, the conditions at float altitudes, which can be reached by zero pressure balloons, could be more stable, enabling accurate pointings. We have used the Inertial Measurement Unit (IMU), placed on a stratospheric zero pressure balloon, to observe 3-axis motion of a balloon payload over a flight 
time of $\sim 4.5$ hours, from launch to the float altitude of 31.2 km. The balloon was launched under nominal atmospheric conditions on May 8th 2016 from a Tata Institute of Fundamental Research Balloon Facility, Hyderabad. 
 
\end{abstract}


\section{Introduction}

High-altitude balloon platforms are an economical alternative to space missions for testing instruments as well as for specific classes of observations, particularly those that require a rapid response such as comets, or other transients. Telescopic platforms at high altitudes have significant advantages over operations from the ground enabling observations at forbidden wavelengths. A UV telescope (200--400 nm) in stratosphere with aperture of just 6 inch in diameter with sufficient pointing stability/accuracy and a 1K$\times 1$K CCD array could provide wide-field images with FWHM better than $1^{\as}$ approaching the diffraction limit (Fesen \& Brown, 2015), similar to that of space observatories but at a much lower cost. We have initiated a high-altitude balloon program at the Indian Institute of Astrophysics to develop low-cost instruments for use in atmospheric and astronomical studies (Nayak et al. 2013, Safonova et al. 2016). We have developed a number of payloads which operate in the near-ultraviolet (NUV), but we are limited to weights under 6 kg for regulatory reasons which constrains our payload size (Sreejith et al. 2016a). Our first experiments were of atmospheric lines (Sreejith et al. 2016b) where the pointing stability is less important, but we plan to observe astronomical sources for which a pointing mechanism is required. 

Unlike in space missions, in stratospheric balloon pointing systems the payload is attached to the balloon by a long flight train. Besides transmitting the balloon's buoyant force, the flight train is the source of disturbances that the pointing control must reject. A typical stratospheric balloon has a train of several meters in length comprised of a recovery parachute beneath with flight termination systems, which is connected to a gondola consisting of scientific equipment with communication and associated electronics. The balloon flight, planned by the Tata Institute of Fundamental Research Balloon Facility (TIFR-BF) on 6--8th May 2016, provided a timely opportunity to piggyback attitude instrumentation on the gondola solely to measure its natural motion during the float phase of the flight. We expect that this information would be helpful in efforts to characterize disturbances that could be expected on any balloon-borne pointing system.

\section{Pointing Systems and Pointing Disturbances on High-Altitude Balloons}

With a telescope placed on the high-altitude balloon payload, one of the major factor in the design of the pointing system is the transfer of oscillations from the flight train; this constraint is more important if the required stability for observation is within a degree of accuracy. In our first experiments of atmospheric lines (Sreejith et al. 2016) the pointing accuracy was not of a great concern, but we plan to observe astronomical sources for which pointing and stabilization of the order of arcseconds are required. The disturbances can excite 
the balloon--payload train dynamics, which at  
the float altitude may disrupt pointing. There are two categories of disturbances:
transverse and azimuth. Transverse disturbances include swinging and bouncing 
oscillations that mostly affect gondola (or telescope) motion  about the two horizontal axes
(elevation and tilt). Azimuth disturbances are 
rotation disturbances about the vertical that similarly affect the telescope azimuth
positioning. 

Payload teams devise methods for mitigating the effects of these disturbances. The
primary strategy is to place the balloon in as quiet an environment as possible during the 
observation, resulting in the so-called `observational windows' --- a period when the winds in 
stratosphere are low and steady (see e.g. Manchanda et al. 2011). In addition, the balloon 
team at TIFR-BF uses the bifilar load line suspension to mitigate the disturbances 
(Robbins \& Martone, 1991).
 
The second strategy is in the careful design of the pointing system. The most common pointing configuration is the azimuth/elevation configuration, where the telescope is tilted about a horizontal elevation axis, fixed in the payload. We have designed and developed a low-cost lightweight, closed-loop pointing system build completely from off-the-shelf components (Nirmal et al. 2016). The system performance was checked on the ground and in tethered flights with satisfactory results. The system can point to an accuracy of $0.1^{\circ}$, and track objects from the ground with an accuracy of $\pm 0.15^{\circ}$.  

It is not certain whether the balloon itself is going to rotate during a flight; however, balloon rotation even during float has been noted in the past (e.g. Gruner et al. 2005).

\section{Instrumentation}

\begin{figure}[h!]
\centering
 \includegraphics[width=.3\textwidth]{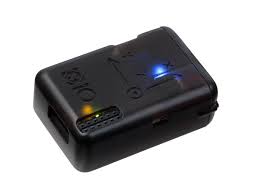}
 \hskip 0.2in
\includegraphics[width=.3\textwidth]{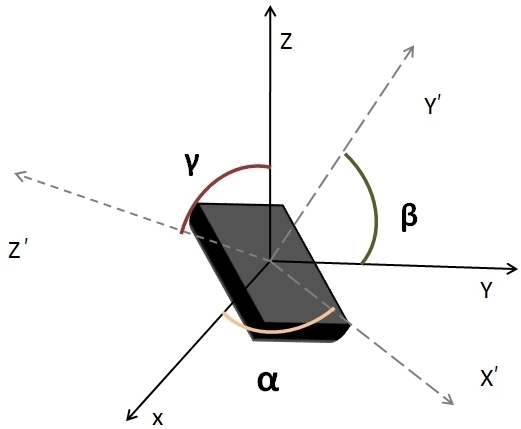} 
\caption{\label{figure:IMU_photo} {\it Left}: The x-IMU with housing and battery. 
{\it Right}: The x-IMU reference frame. Axes $X$, $Y$ and $Z$ define Earth-centered inertial (ECI)  reference 
frame; $X'$, $Y'$ and $Z'$ define body-centered reference frame, and angles 
$\alpha$, $\beta$ and $\gamma$ are the Euler angles.}
\end{figure}

\begin{figure}[h!]
\centering
 \includegraphics[width=.25\textwidth]{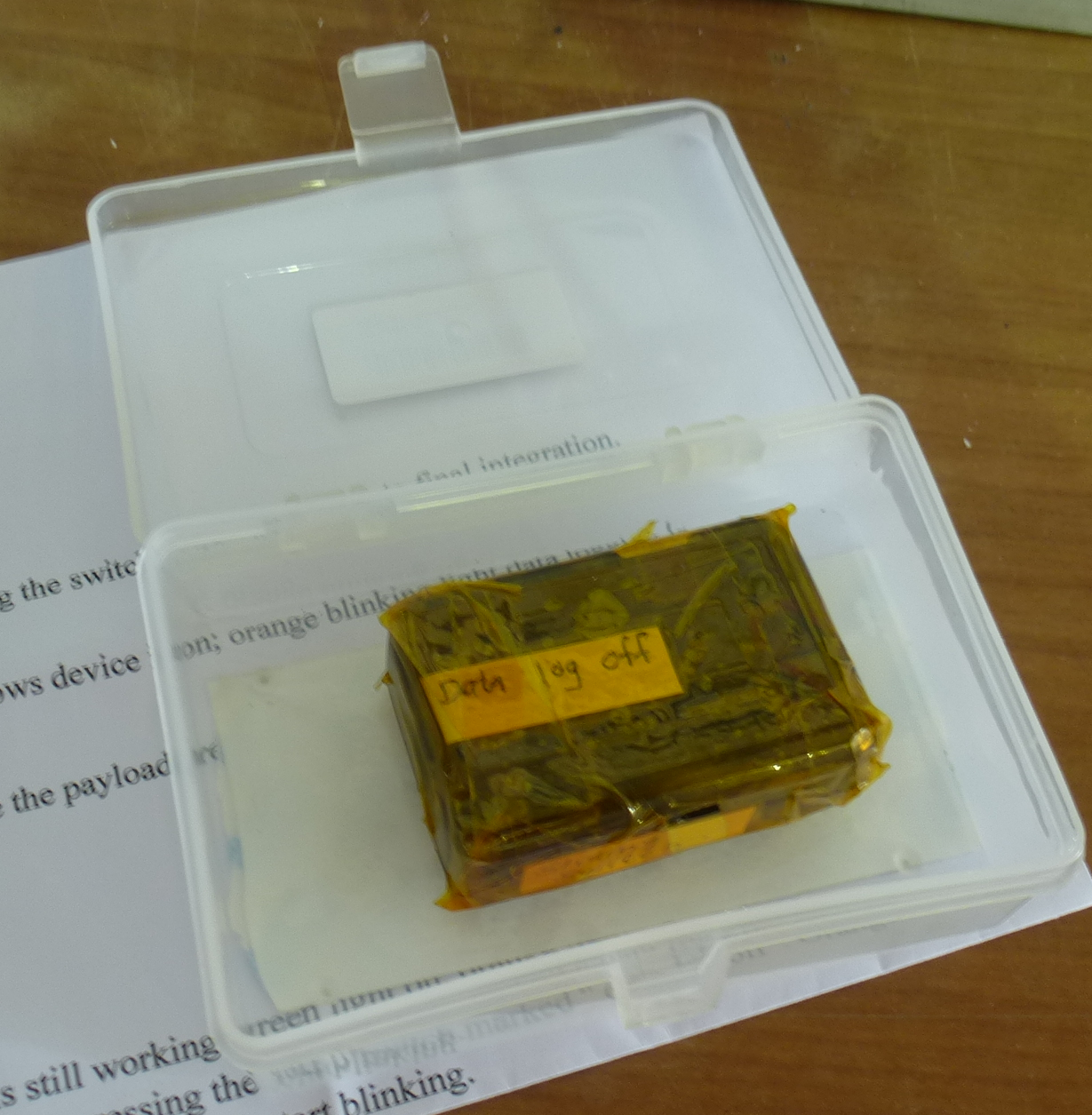}
\includegraphics[width=.32\textwidth]{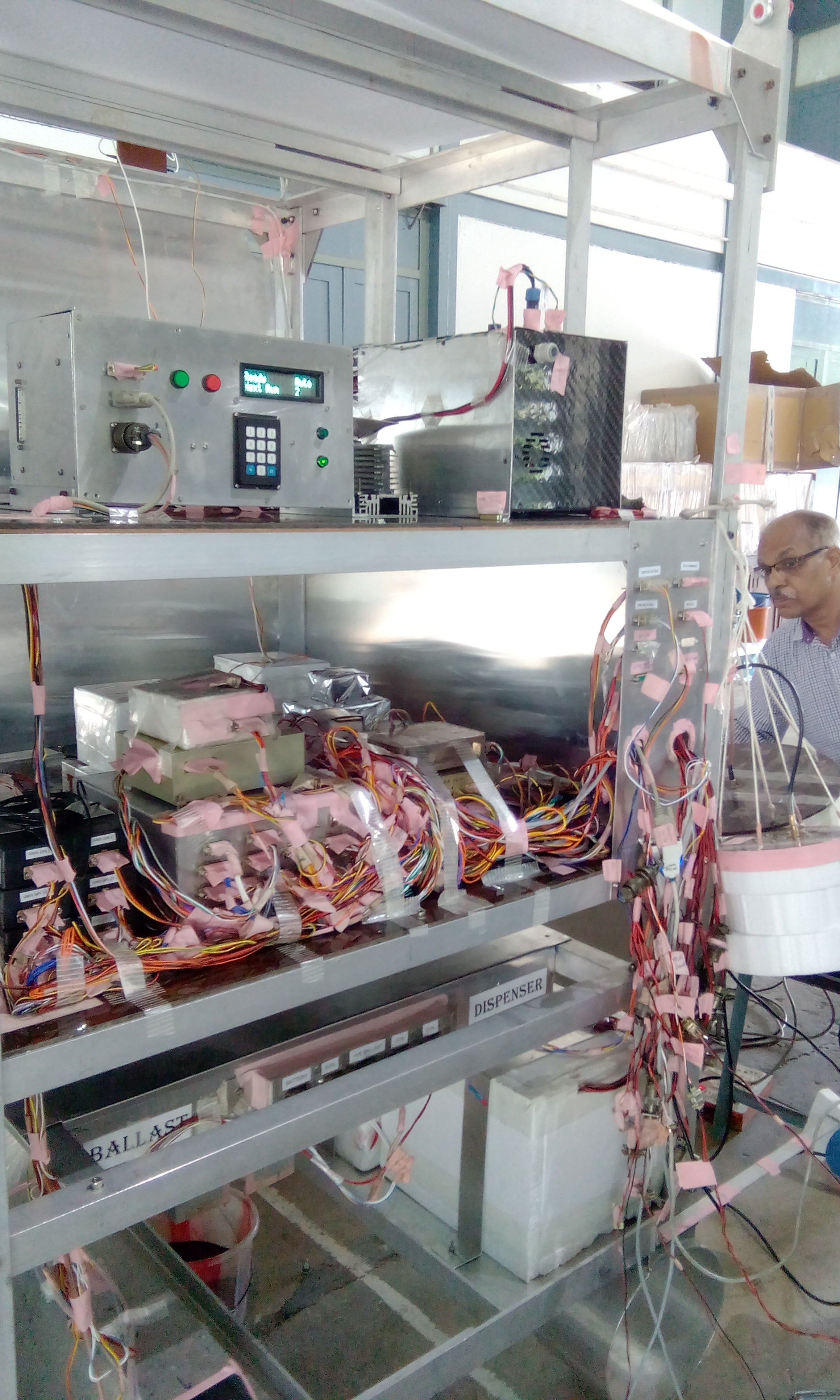}
\includegraphics[width=.37\textwidth]{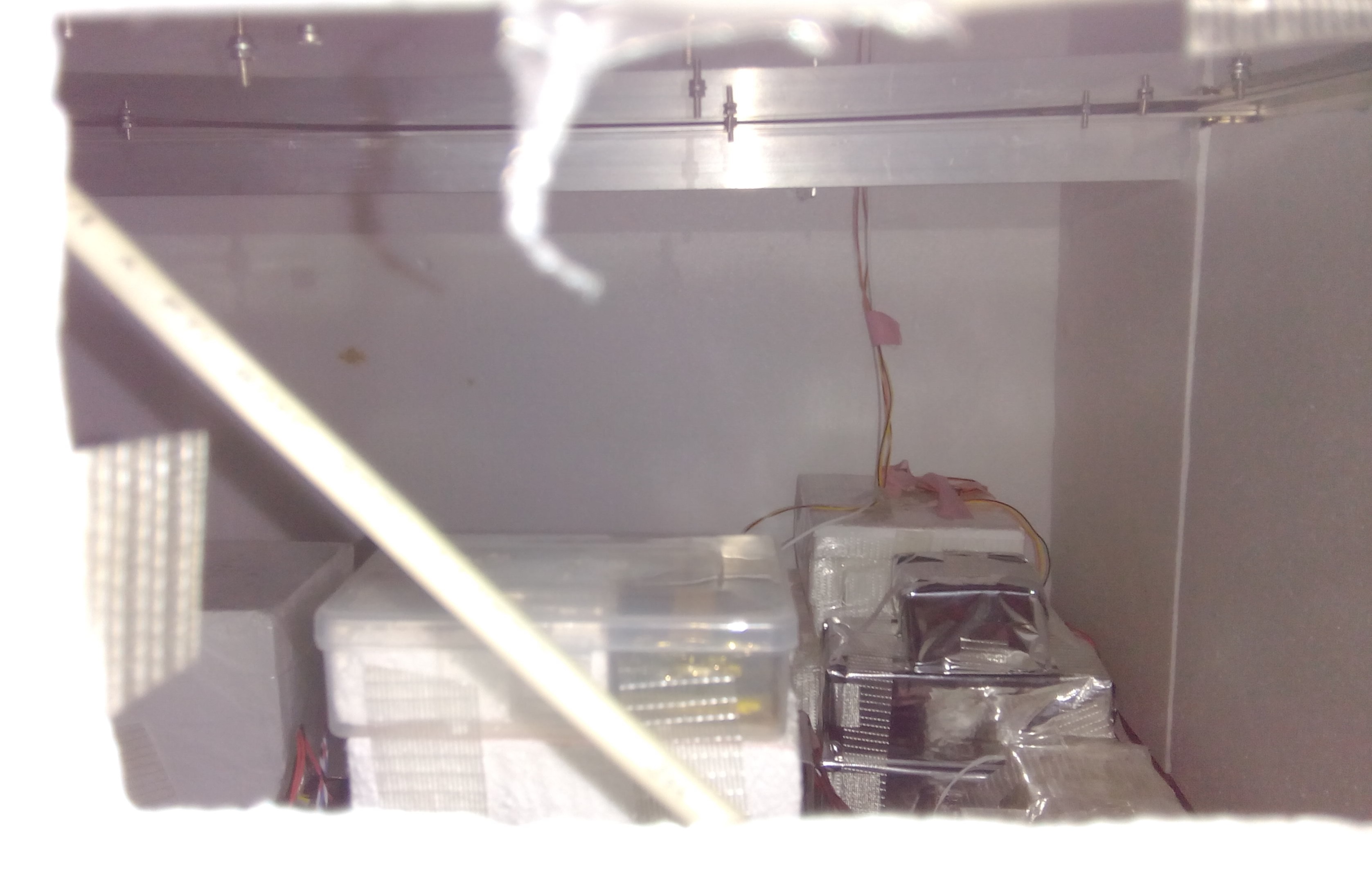}
\caption{\label{figure:IMU_payload} {\it Left}: x-IMU before integration with the main 
payload. {\it Middle}: 
The main payload before closing the panels. The IMU will be placed in the middle tier. 
{\it Right}: Final placement inside the main payload. A view through the specially made cut-out in the front panel.}
\end{figure}

With no requirement to point the gondola on this flight, an inertial measurement unit (IMU)\footnote{From x-io Technologies, UK, http://www.x-io.co.uk/products/x-imu/} with 
internal Li-polymer battery (tested previously in the lab to work for up to 12 hrs) and a 
4 GB SD card for data logging (Fig.~\ref{figure:IMU_photo} and Table~\ref{table:imu}), 
was placed into the main payload (gondola) (Fig.~\ref{figure:IMU_payload}).

\begin{table}[ht!]
\begin{center}
\caption{Technical specifications of the X-IMU sensor} 
\begin{tabular}{lc} 
\hline
Dimensions & $57\times 38 \times 21$ mm \\
Weight & 49 gms \\
Operating Temperature & -30 C to +85 C \\
Power & internal Li-polymer battery 3.7 V 1000 mAh \\
Data storage & internal 4GB SD card \\
Components   & 3-axis accelerometer, 3-axis gyroscope, 3-axis magnetometer, T sensor\\
\hline
\end{tabular}
\label{table:imu}
\end{center}
\end{table}

The x-IMU uses Euler angle transformation to calculate yaw $\alpha$, pitch $\beta$  
and roll $\gamma$ angles (Fig.~1). The roll and pitch angles are derived from the 
accelerometer and gyroscope output, and the yaw readings from the magnetometer. According 
to the way we place the x-IMU in the payload, the yaw angle is equivalent to the azimuth, 
pitch is equivalent to the elevation, and tilt is equivalent to the roll angle. Since we 
are only interested in the motion of the payload, not the absolute position, we take these 
values of roll, pitch and yaw straight away for azimuth, elevation and tilt without any conversion. 

\section{Flight Description}

\begin{figure}[h!]
\centering
 \includegraphics[width=.6\textwidth]{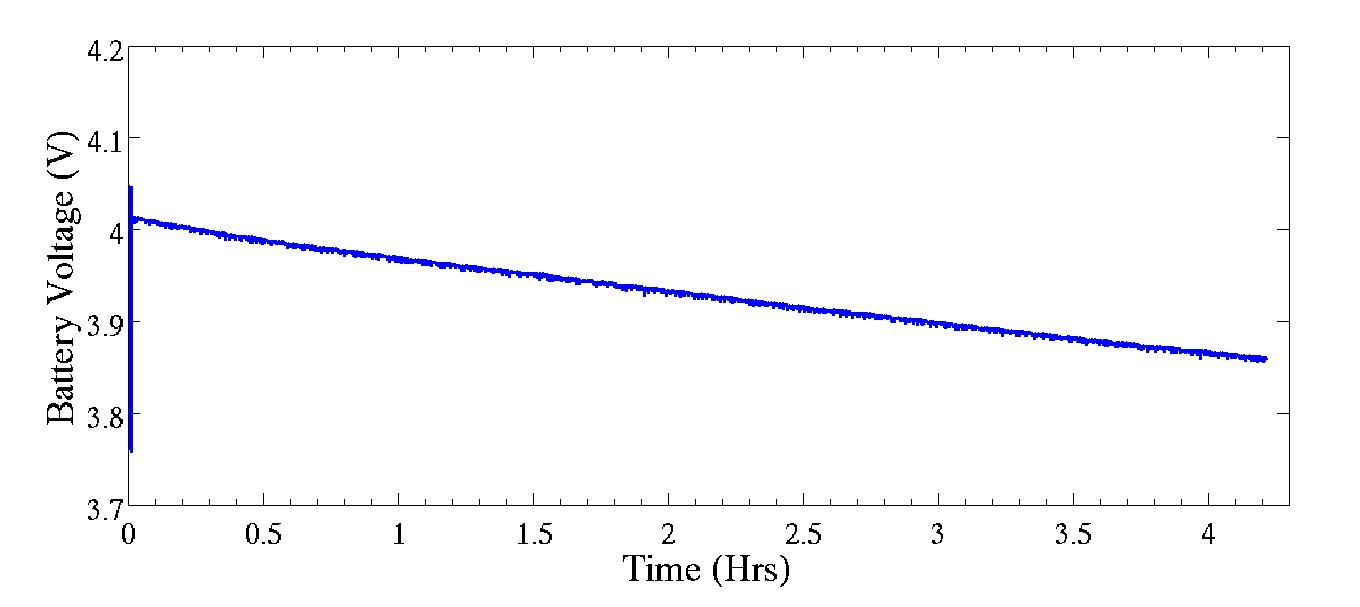}
\caption{Battery voltage variation with time.}
\label{fig:BAT} 
\end{figure}

The initial launch was proposed to be conducted before sunrise on May 6th 2016, however, weather conditions deteriorated rapidly and delayed the launch till the next day. The balloon was launched on May 7th 2016 at 6:43 am from the TIFR-BF, Hyderabad, 
India (17.4729N, 78.5785E, altitude 532 m). The weather conditions were fairly clear with thin
clouds and favourable surface winds. Several rubber 
weather balloons were tethered-launched prior to the mission to test the direction and strength of the  
surface winds. The IMU was switched on at 5:35 am, few minutes before the final integration with 
the main payload. Total payload weight was $\sim 200$ kg, connected to the balloon through the bifilar suspension. The main balloon was 20 $\mu$m-thick 
plastic zero-pressure balloon of volume 38,211 $m^3$ filled with hydrogen. The average ascent rate was 4.59 m/sec. 
The balloon reached the float altitude of 31.2 km at 08:36 am (ascent time 1 hr 53 min). The flight was 
terminated by the onboard programmable timer at 12:42 pm IST on May 7th, and the payload with deployed parachute landed 380 km west of Hyderabad in good condition. The IMU was recovered on May 7th at 4:00 pm, with the power 
and data logging still on at the landing site. The IMU was 
powered off without switching off the data logging, and delivered to the TIFR-BF at 11:30 am on May 8th. In 
order to keep the data intact, the IMU was powered on and the data 
logger was switched off, followed by the IMU power off. The battery was working continuously for nearly 11 hrs, and  
there was still power in the battery even after 22 hrs. In Fig.~\ref{fig:BAT}, we show the trend of the voltage with time for the first $\sim 4.5$ hrs of the flight.

\section{Observations and Results}

We have combined the data from our x-IMU and the data from measurements by the TIFR BF from the main payload. Because the x-IMU was powered off 
before turning off the data logging, a lot of data got corrupted. We have recovered the data from only the first 
$\sim 4.5$ hours of flight ($\sim 2.4$ hrs of float), which is sufficient for our purposes. 

\subsection{Stratospheric Winds}

\begin{figure}[ht!]
\hspace{-0.5in}
\includegraphics[scale=0.56]{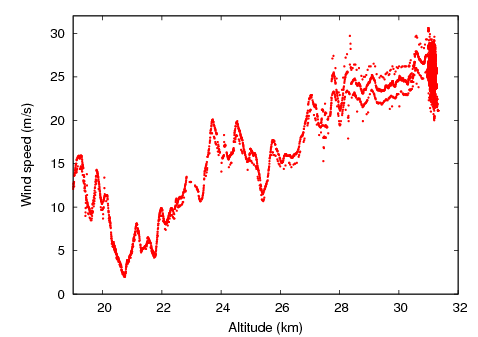}
\hspace{-0.18in}
\includegraphics[scale=0.56]{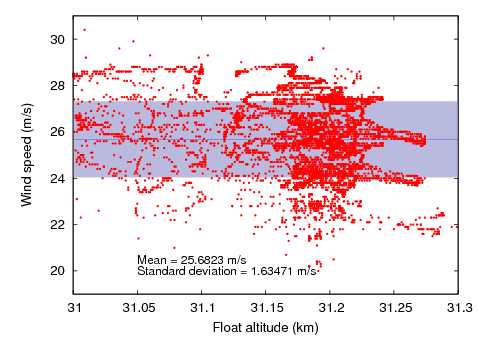}
\caption{Wind speed in stratosphere ({\it Left}) and at float ({\it Right}) from the TIFR-BF data. Shaded area shows standard deviation about the mean.}
\label{figure:windspeed} 
\end{figure}

Despite the lack of atmosphere ($\lesssim 3$ mbar), the stratosphere both directly and indirectly affects the balloon-borne payload via stratospheric wind shears, causing translational and rotational acceleration as well as rotation.  
The speed and direction of stratospheric winds obtained by iMet-1 GPS Radiosonde\footnote{InterMet Systems, Inc. South Africa. {\tt http://http://intermetsystems.com/}.}, were provided by the TIFR Balloon Facility. We need the speed of the wind to estimate its influence on the motion of the payload, and compare the stratospheric wind conditions at float with the near-surface conditions. The wind speed in stratosphere is shown in Fig.~\ref{figure:windspeed}.

It is interesting to note that despite density decreasing with altitude, the lowest winds were reported at altitudes of about 20 km, rising after that (Fesen \& Brown, 2015). We also noticed that at $\sim 20$ km, the wind speeds were the least, at only 2--5 m/sec, while at the float altitude, $\sim 31$ km, the mean speed was higher at $\sim 25.5$ m/sec.

\subsection{Temperature variation}

\begin{figure}[h!]
\centering
\includegraphics[scale=0.27]{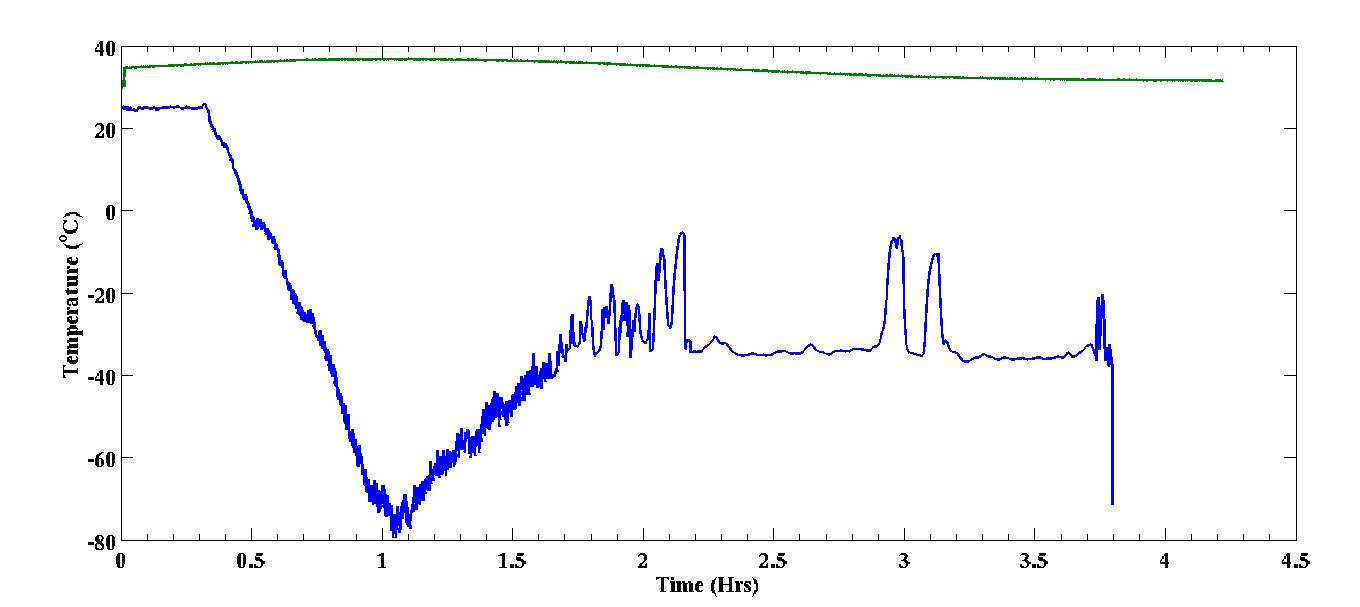}
\caption{\label{figure:temp} Temperature outside (blue line, TIFR data) and inside the payload box (green, IMU data).}
\end{figure}

Figure~\ref{figure:temp} shows the temperature inside (measured by our IMU) and outside 
(measured by the TIFR-BF) the payload box for the first $\sim 4.5$ hours of flight.

\subsection{Payload Motion}

\begin{figure}[ht!]
\centering
\includegraphics[width=0.8\textwidth]{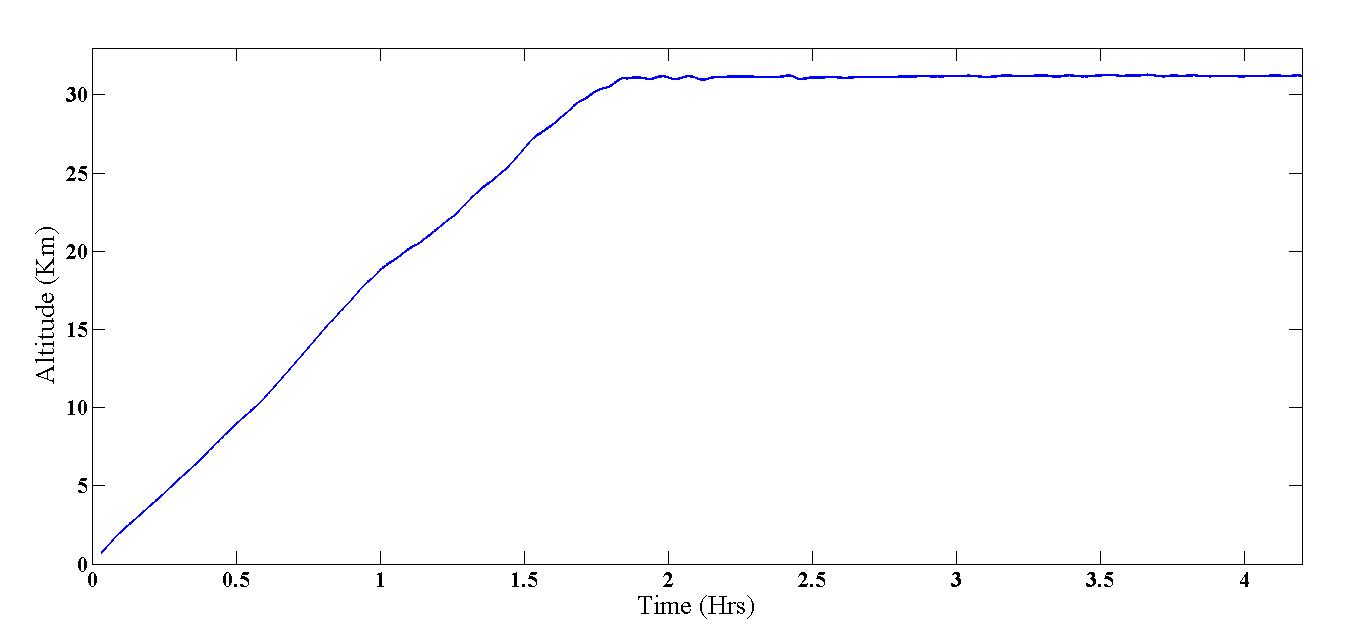}
\caption{\label{figure:height} The variation of altitude with time from launch, obtained by the TIFR-BF.}
\end{figure}

The variation of altitude with time for the first $\sim 4.5$ hours of the flight is shown in Fig.~\ref{figure:height}.

There was no significant surface wind at launch, however, the disturbance from the load train caused the payload to 
swing and rotate at the ascent (Fig.~\ref{figure:ascent}). The IMU provides data on the payload motion 
in 3-axes. In addition, we can extract the accelerations in 3 directions. 

\begin{figure}[ht!]
\centering
 \includegraphics[scale=0.173]{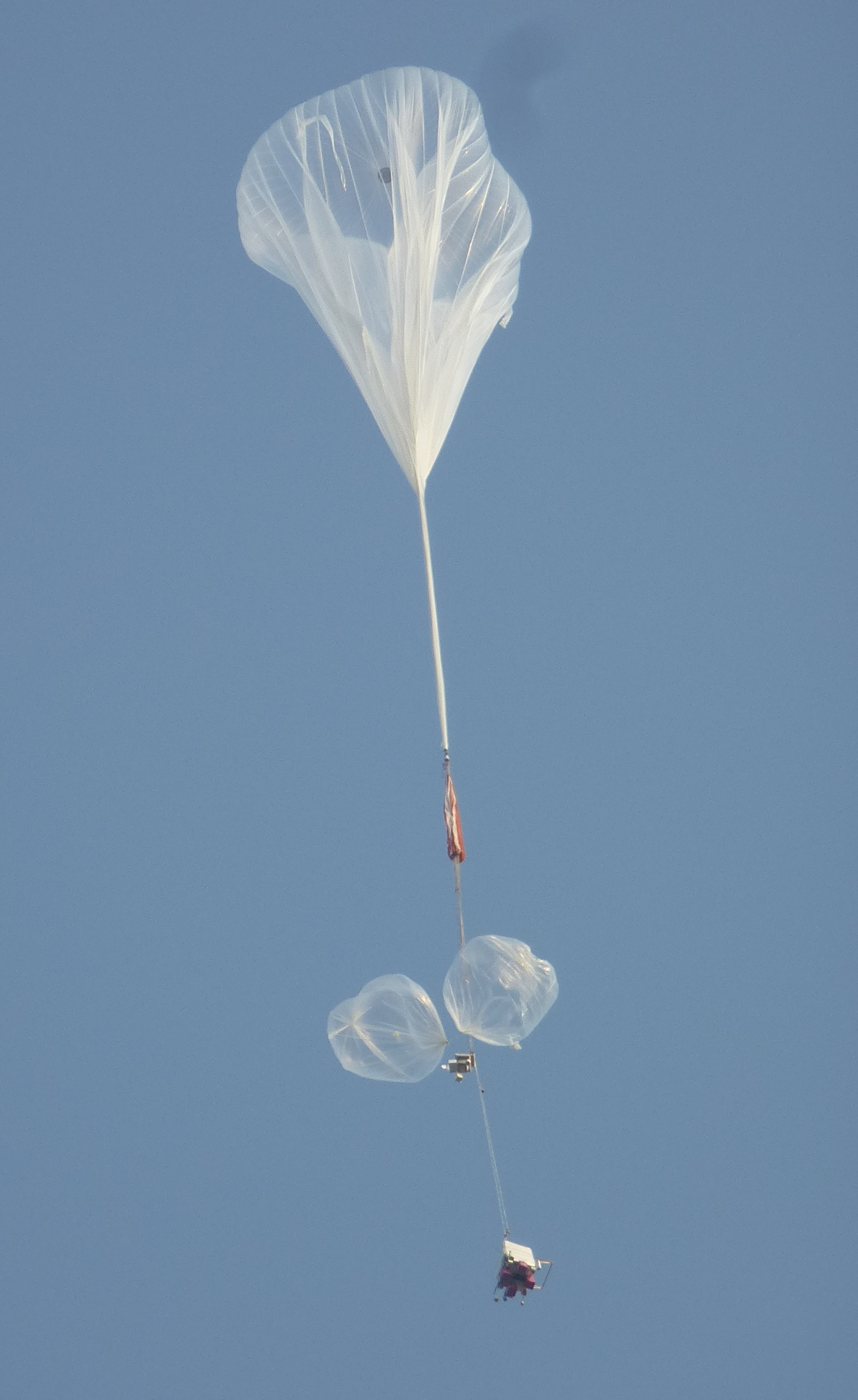}
\includegraphics[scale=0.29]{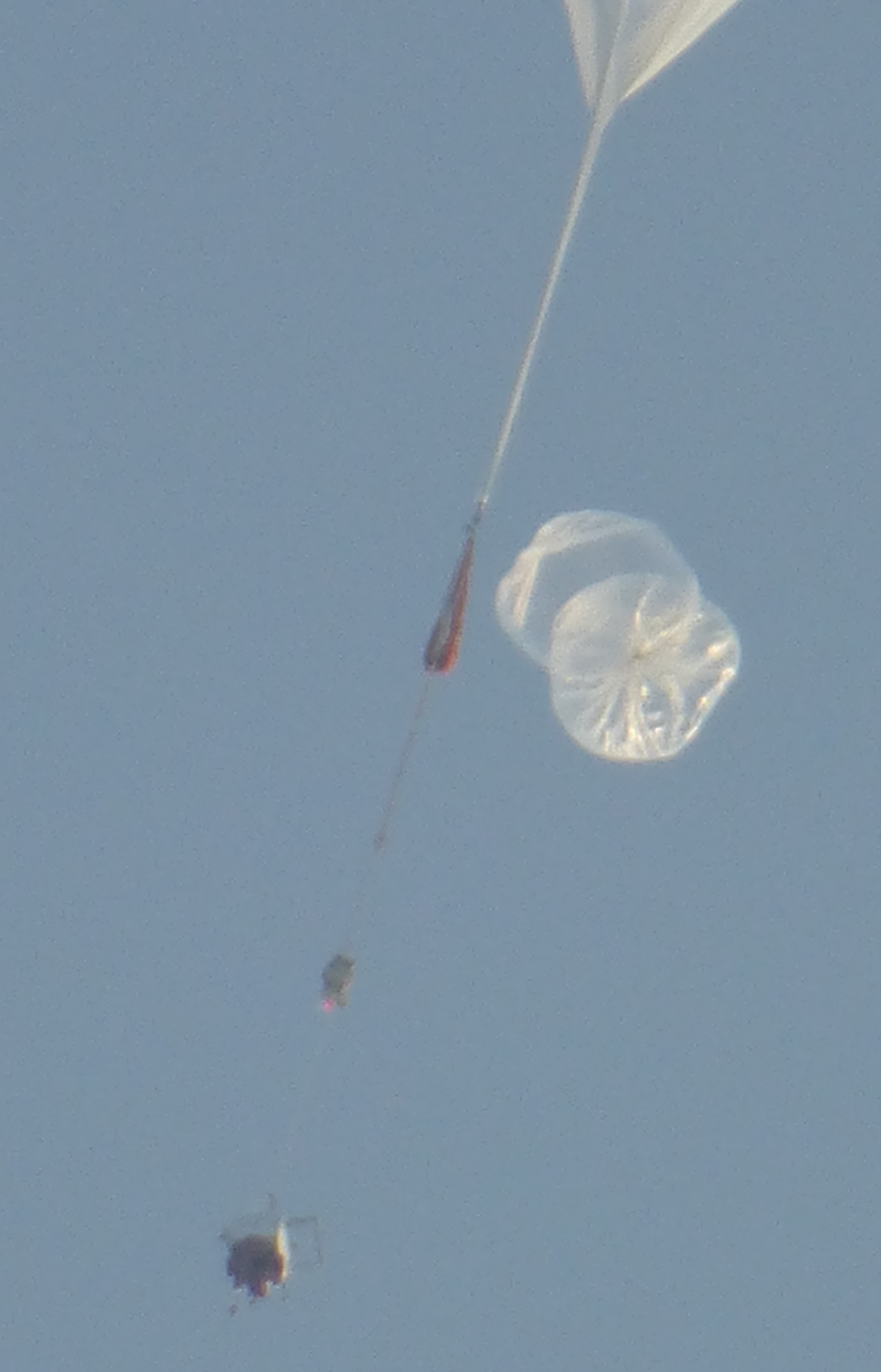}
\caption{\label{figure:ascent} Swinging and rotation of the payload in the first minutes of ascent.}
\end{figure}

\begin{figure}[hb!]
\vspace{0.3in}
\centering
\hspace{-0.35in}
\includegraphics[scale=0.57]{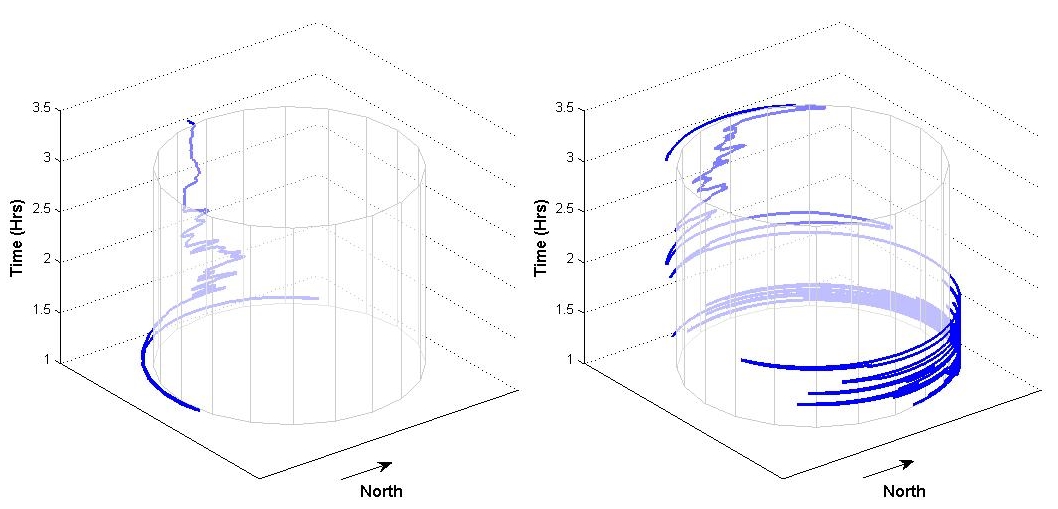}
\caption{\label{figure:ascent} Wind direction ({\it Left}) and movement of the payload in azimuth ({\it Right}) after reaching the stratosphere. Time axis shows time in hours elapsed since launch. The float altitude was reached at about 2 hrs after launch. The plots clearly show the difference between the rapid variation of payload azimuth before reaching the final float altitude and less variation of payload azimuth after reaching the float altitude. }
\label{fig:cylinder}
\end{figure}

\begin{figure}[ht!]
\includegraphics[width=0.56\textwidth]{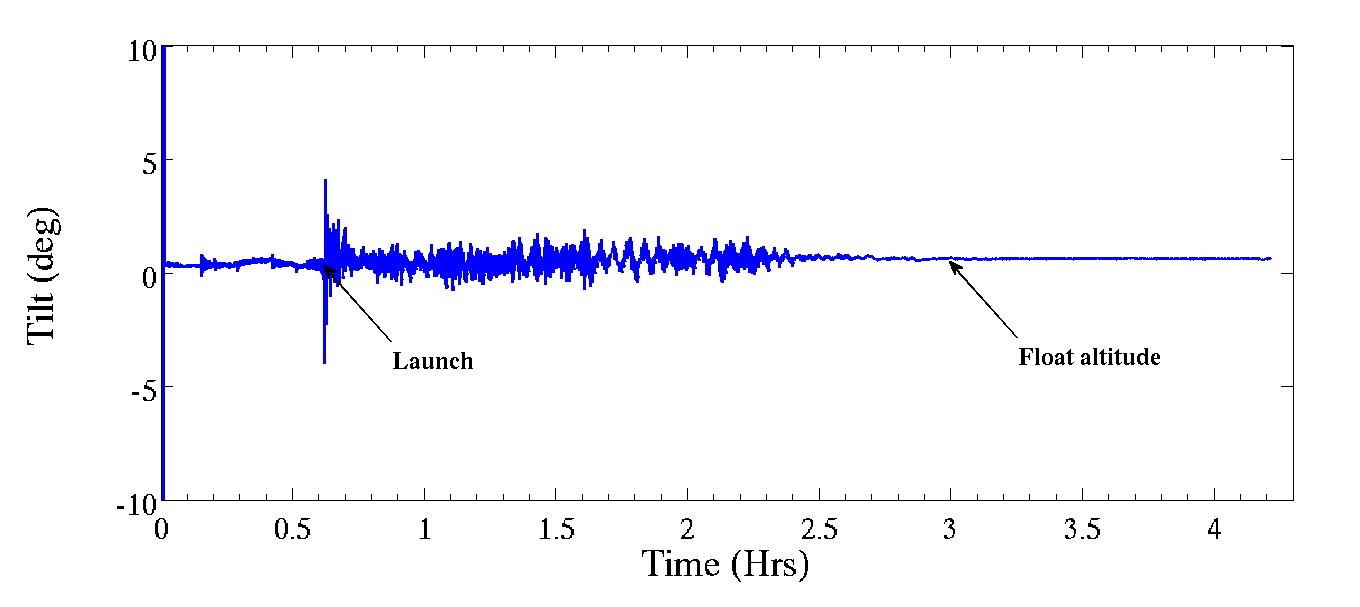} 
\includegraphics[width=0.53\textwidth]{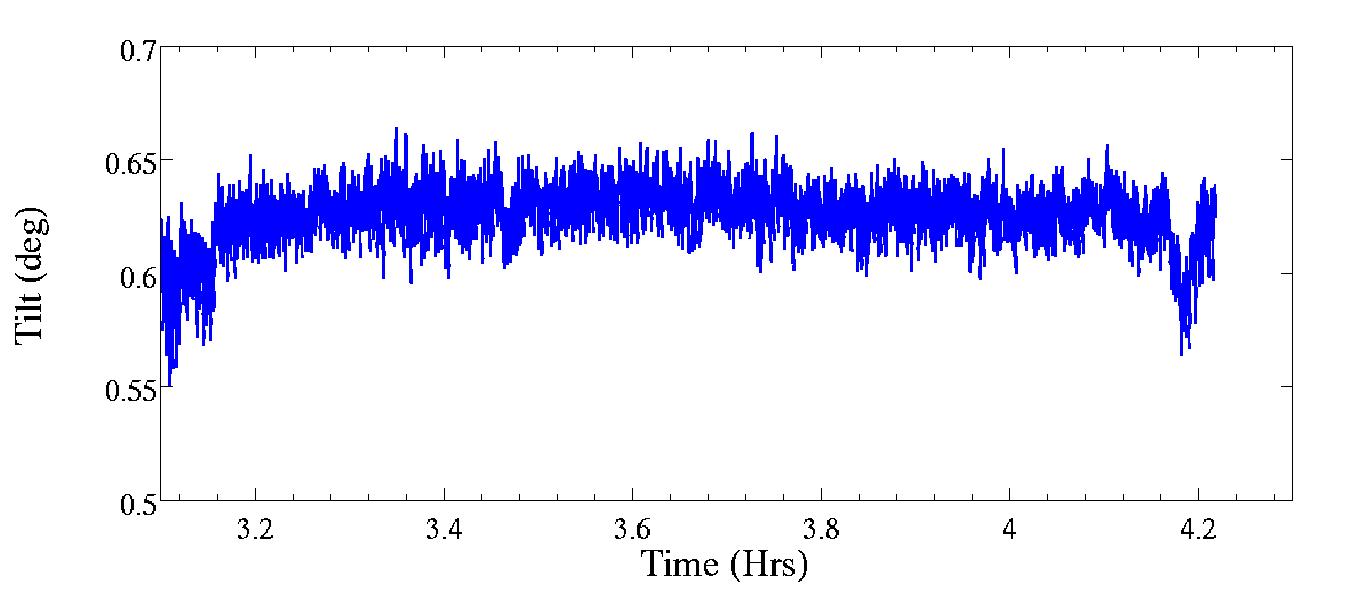} 
\includegraphics[width=0.56\textwidth]{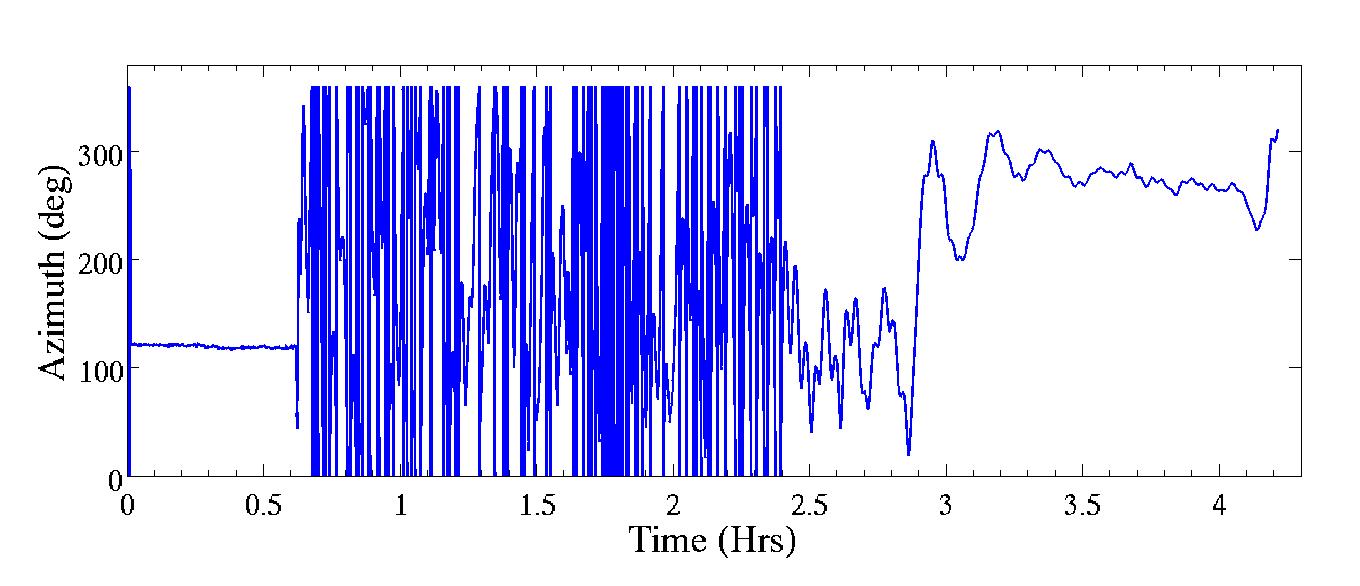} 
\includegraphics[width=0.53\textwidth]{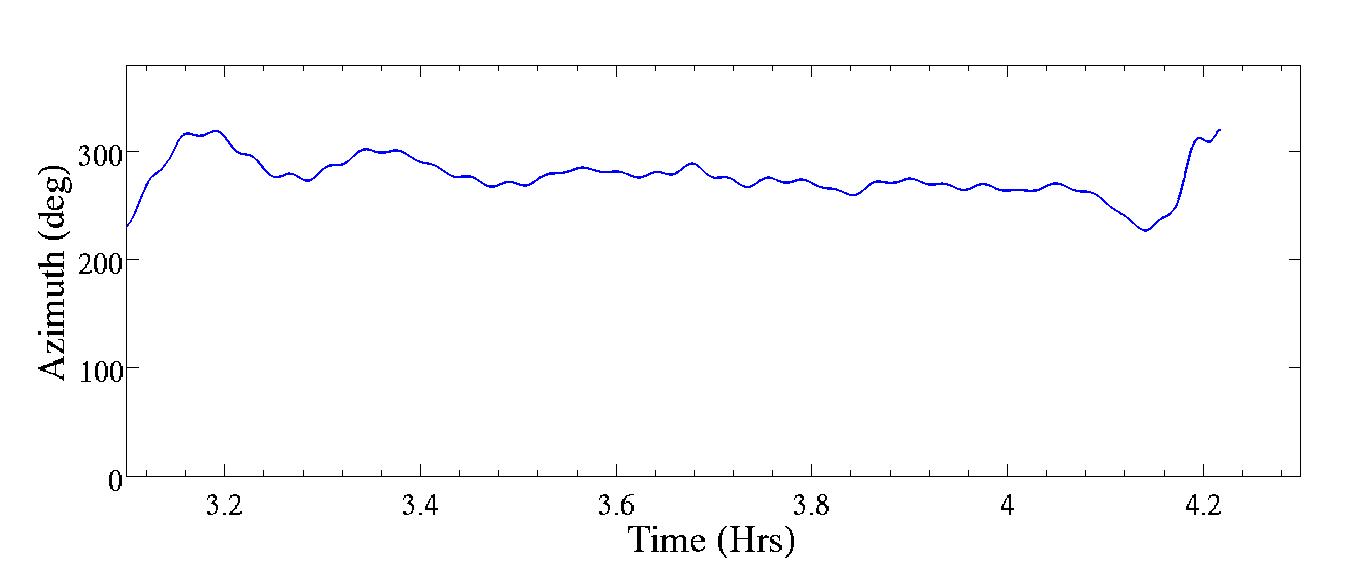} 
\includegraphics[width=0.56\textwidth]{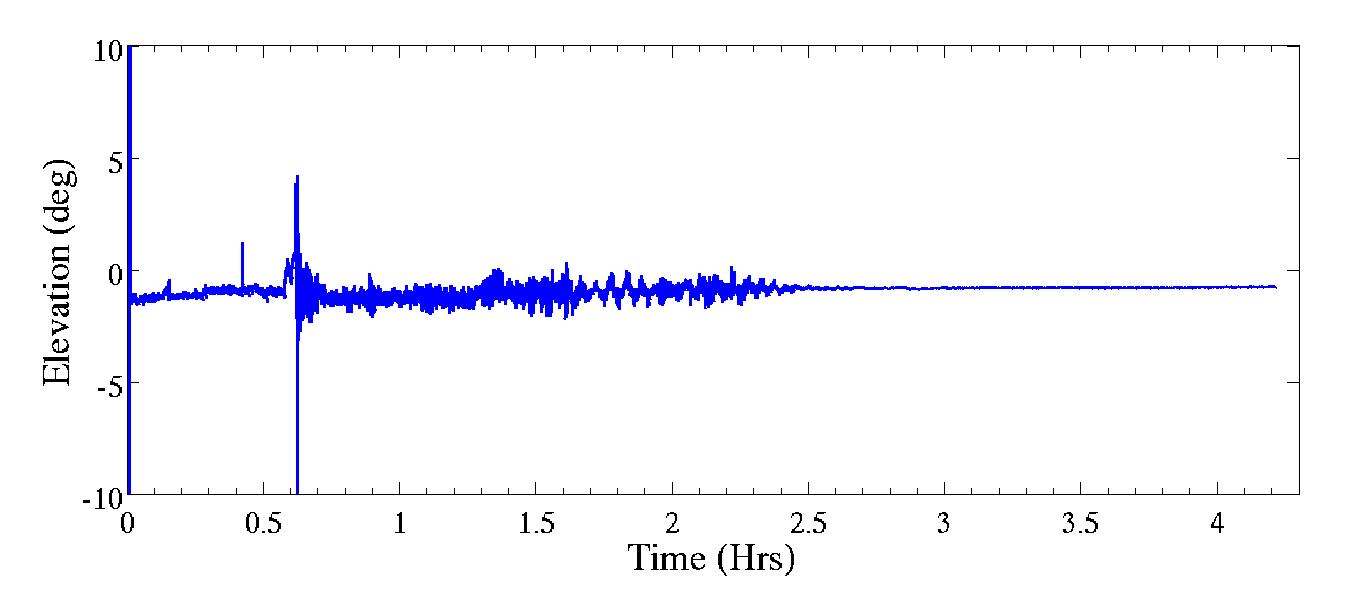} 
\includegraphics[width=0.53\textwidth]{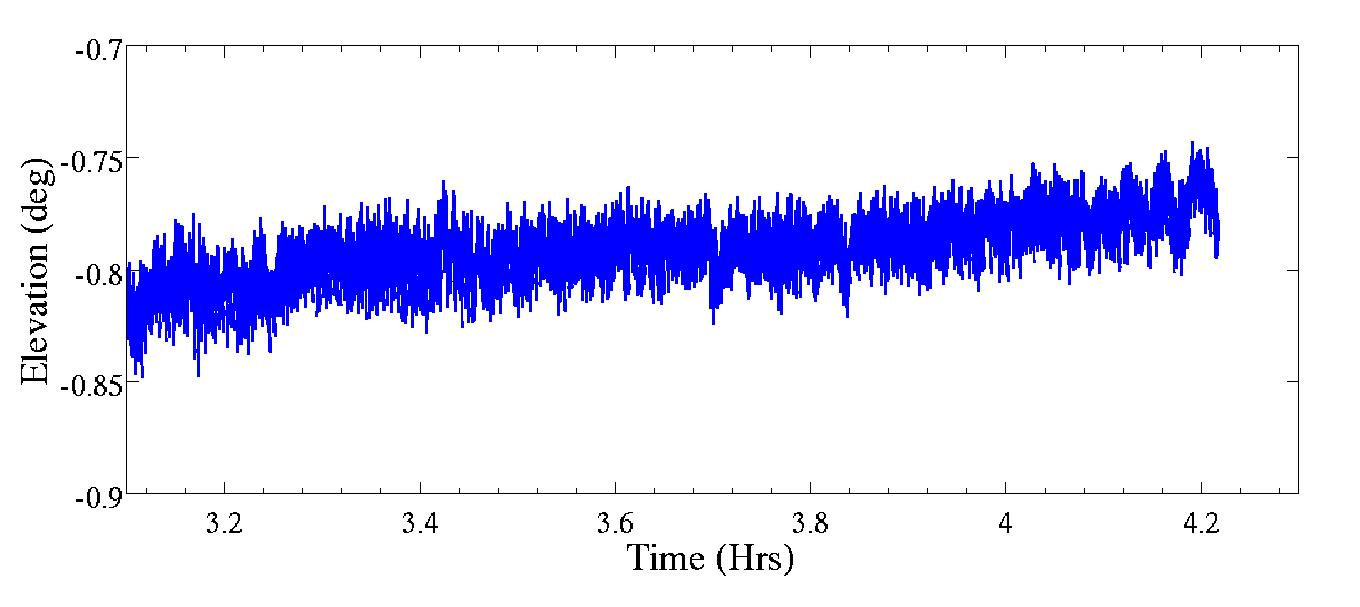} 
\caption{\label{figure:payload_rot} Movement of payload in tilt (roll) ({\it Top}), azimuth (yaw)  
({\it Middle}), and elevation (pitch) ({\it Bottom}). $X$-axis is time in hours from the switch on at 5:35 am IST. 
On the left is the total retrieved data, and on the right is the data at the float altitude.}
\end{figure}

\begin{figure}[hb!]
\centering
\includegraphics[scale=0.17]{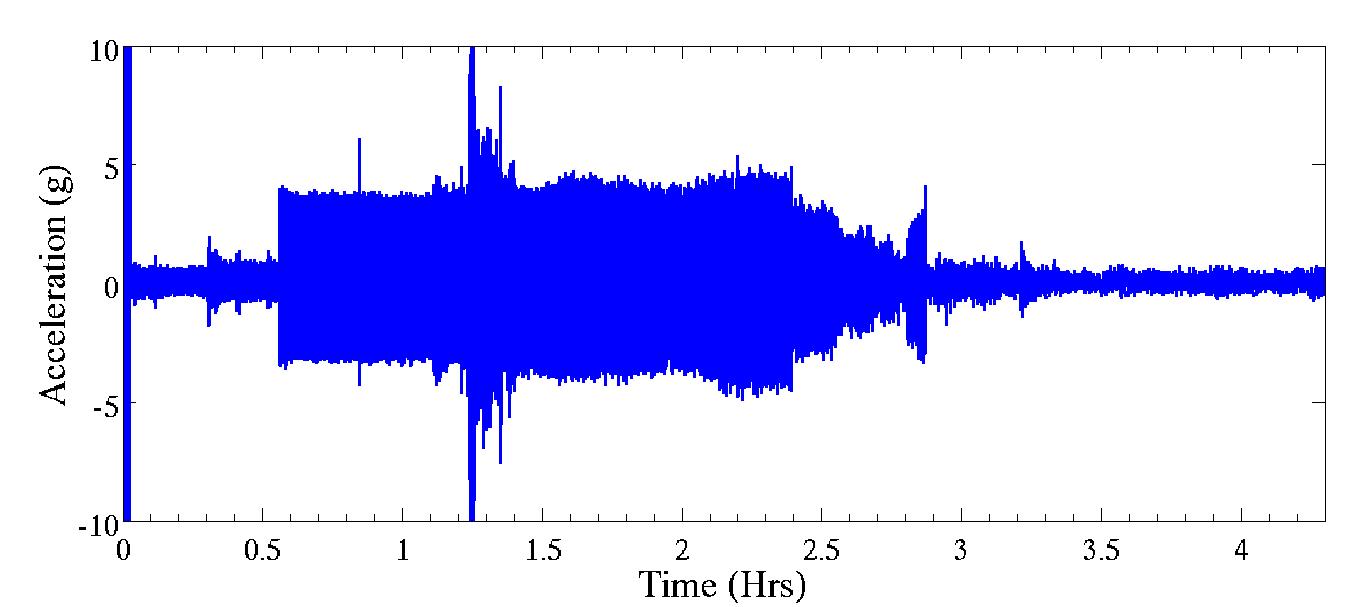}
\includegraphics[scale=0.17]{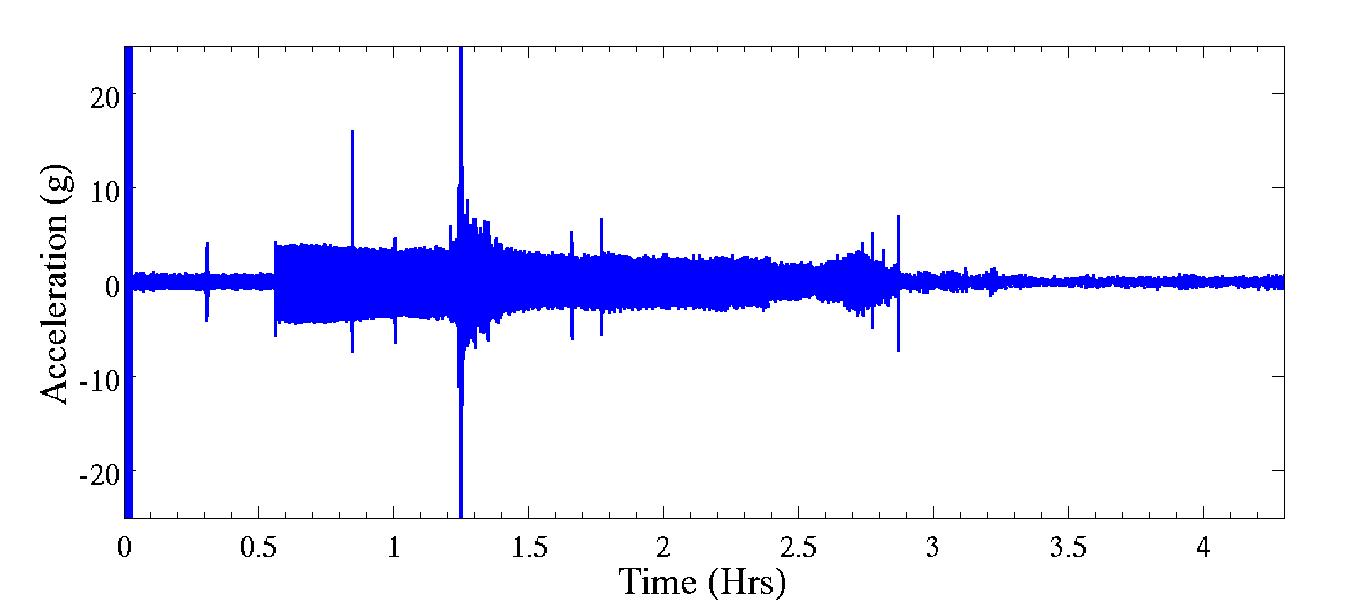}
\hspace{-0.35in}
\includegraphics[scale=0.17]{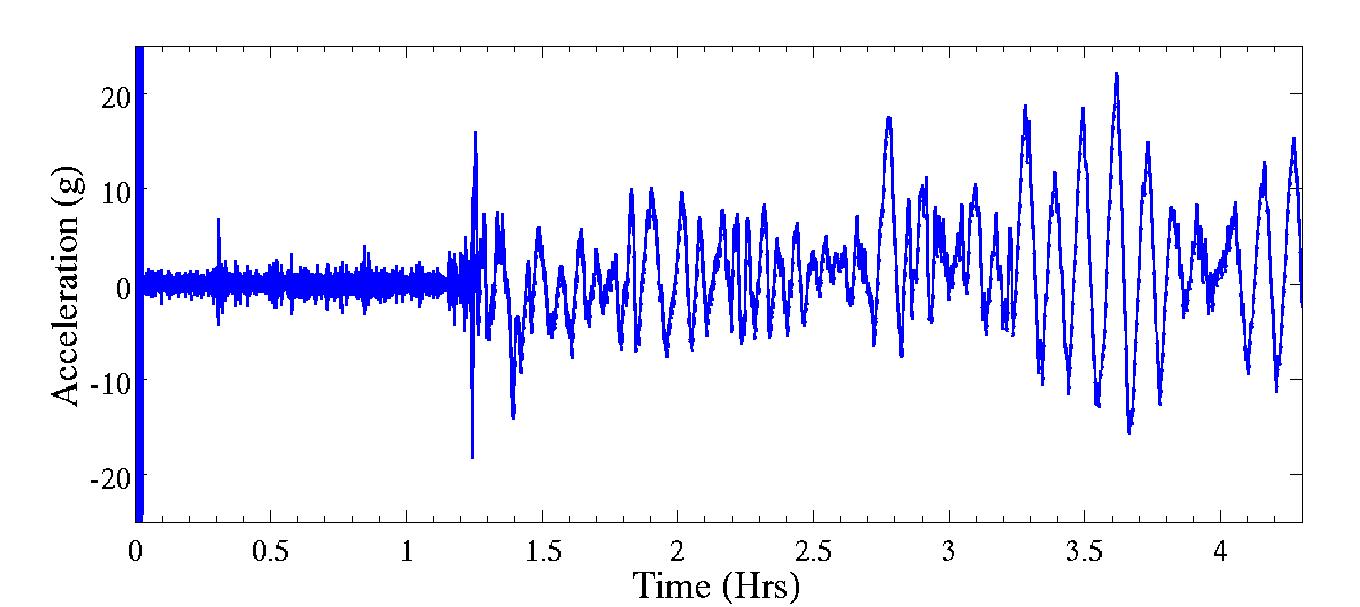}
\caption{\label{figure:acc} The acceleration of payload in $X$ ({\it Top}), $Y$ ({\it Bottom left}) 
and $Z$ ({\it Bottom right}) axes during the flight.}
\end{figure}

The payload motion in roll (tilt), pitch (elevation) and yaw axis (azimuth) measured by our x-IMU is displayed in 
Fig.~\ref{figure:payload_rot}. We also obtained data on payload acceleration in three axes  (displayed in Fig.~\ref{figure:acc}). We have correlated the motion of the payload at float with the changes in the wind direction. This is shown in Fig.~\ref{fig:cylinder}.
The plot on the left shows the change in the direction of the wind with respect to the North with time. The wind direction data points were generated from the GPS coordinates of the main payload with time resolution of 10 seconds. The wind direction plot is in meteorological convention. That is, for example, an angle of $110\deg$ from the North shows that the wind is blowing from $110\deg$ to its diagonally opposite point which is $250\deg$. The plot on the right shows the changes in the payload azimuth (yaw angle). Time axis shows the time from launch: at 1.5 hours the balloon reached an approximate altitude of 25 km, at $\sim 2$ hours the balloon reached the float altitude (see Fig.~\ref{figure:height} for the plot of altitude with time) till the end of available data from the x-IMU (about 3.1 hrs from launch). In this layer of the atmosphere we observe reduced wind turbulence comparing to the lower layers of the atmosphere and the reduction in the disturbances of the payload at float altitude. 

\subsection{Magnetic field}

\begin{figure}[h!]
\centering
\includegraphics[scale=0.17]{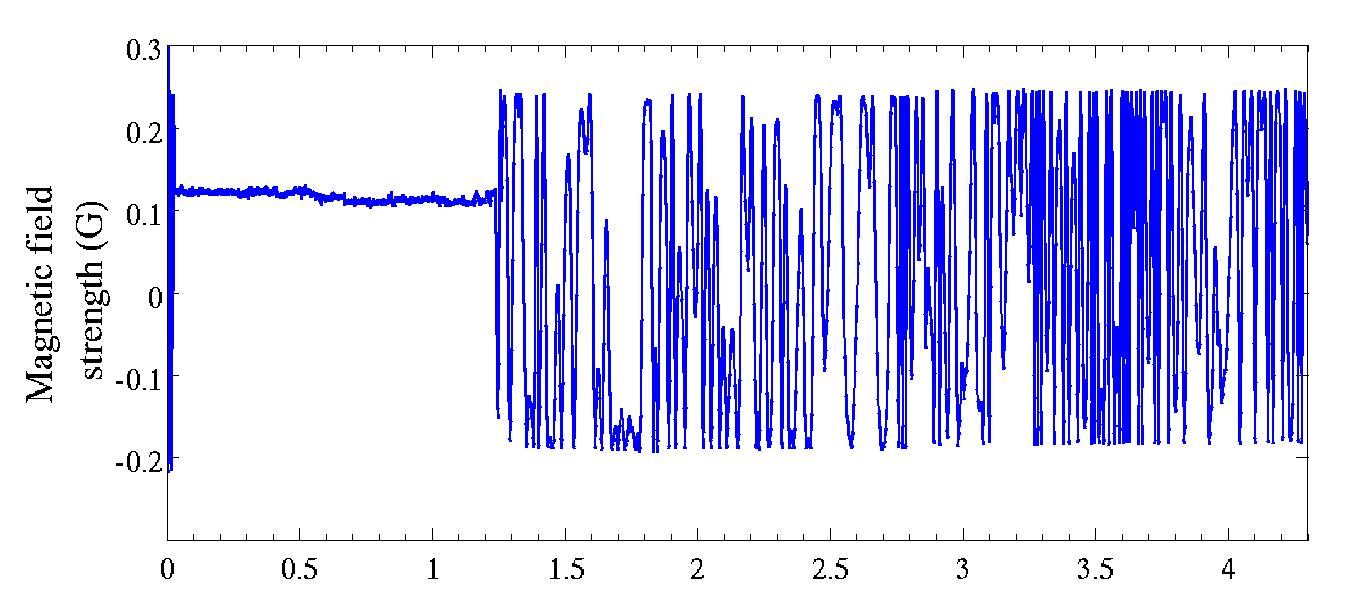}
\includegraphics[scale=0.17]{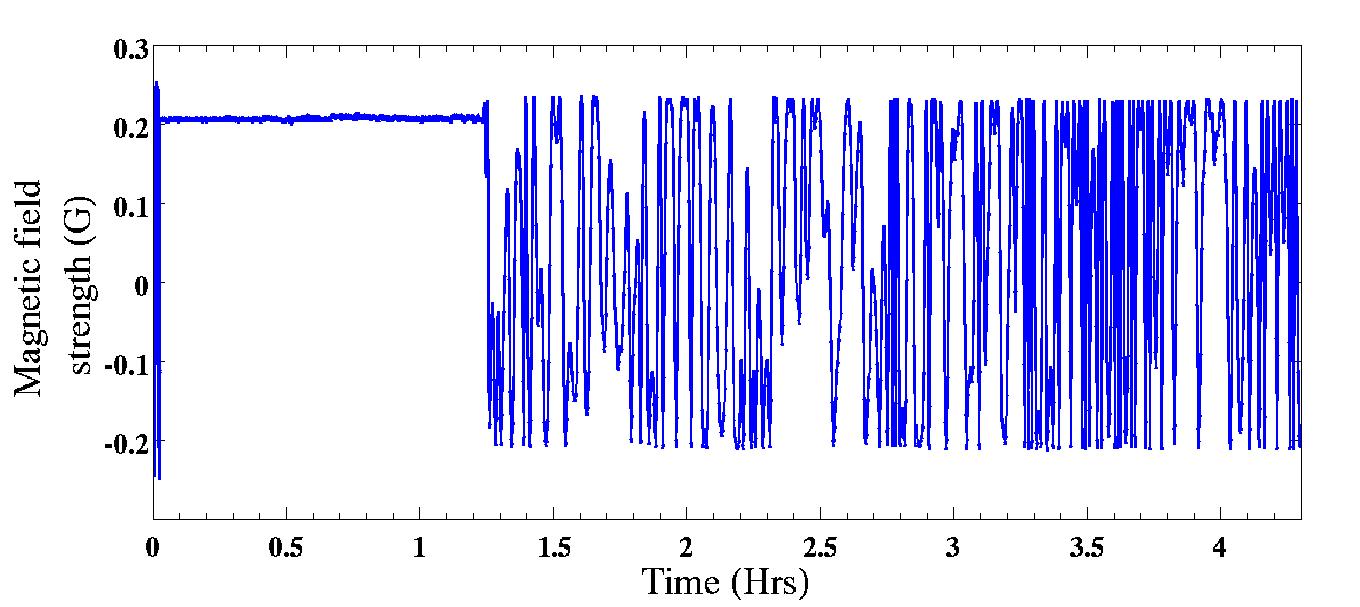}
\hspace{-0.4in}
\includegraphics[scale=0.17]{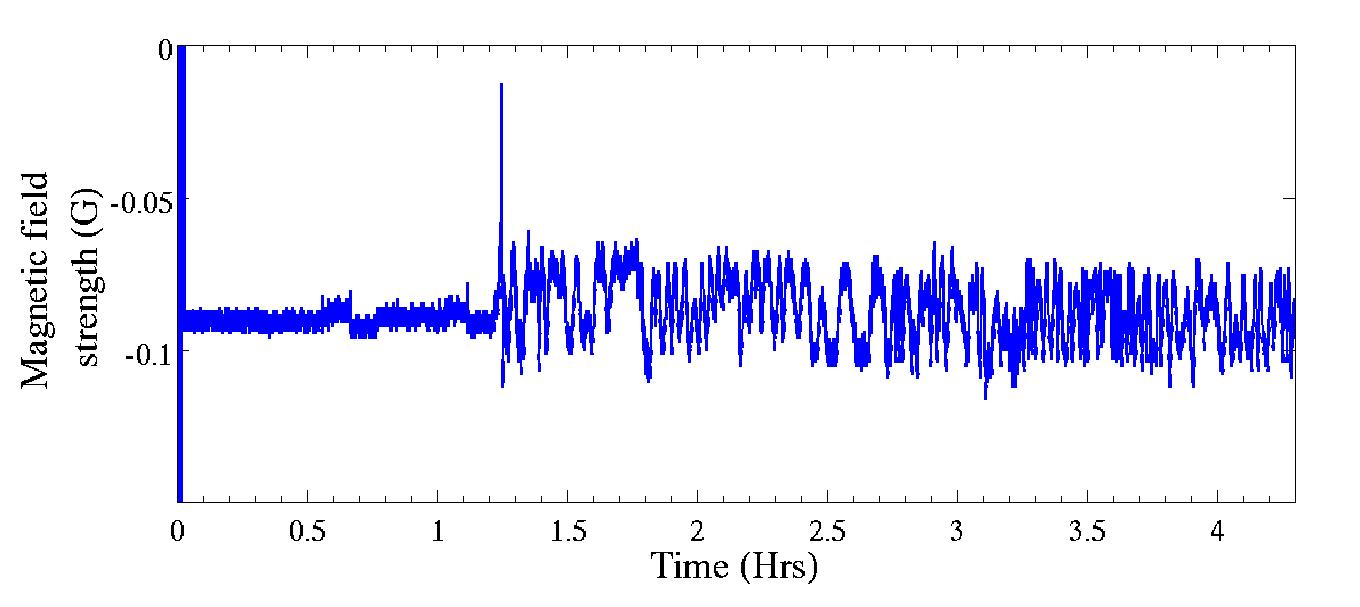}
\caption{\label{figure:mag} The variation of magnetic field strength in $X$ (Top), 
$Y$ (Middle) and $Z$ (Bottom) axes during the flight.}
\end{figure}

The variation of magnetic field around x-IMU in $X,Y,Z$ axes is shown in Fig.~\ref{figure:mag}. 

\section{Summary and Conclusions}

The flight data derived from our x-IMU and from measurements by the TIFR-BF main payload are consolidated in Table~\ref{table: flight_summary}.
\begin{table}[hb!]
\caption{Summary of the flight ($\sim 2.5$ hrs at float altitude)} 
\label{table: flight_summary}
\begin{center}       
\begin{tabular}{|l|c|}
\hline
\rule[-1ex]{0pt}{3.5ex}  Maximum height reached by payload &   31.4 Km \\
\hline
\rule[-1ex]{0pt}{3.5ex} Total duration of the flight &  5 hrs 56 mins \\
\hline
\rule[-1ex]{0pt}{3.5ex} Average ascent rate &  4.59 m/s \\
\hline
\rule[-1ex]{0pt}{3.5ex} Float altitude &  31.2 km \\
\hline
\rule[-1ex]{0pt}{3.5ex} Float reached (time) & 08:35 am IST  \\
\hline
\rule[-1ex]{0pt}{3.5ex} \begin{tabular}{@{}c@{}} RMS motion of payload  \\ during float in azimuth ($^{\circ}/s$) \end{tabular} & 0.5865  \\
\hline
\rule[-1ex]{0pt}{3.5ex} \begin{tabular}{@{}c@{}} RMS motion of payload  \\ during float in elevation ($^{\circ}/s$) \end{tabular} &  0.0104 \\
\hline
\rule[-1ex]{0pt}{3.5ex} \begin{tabular}{@{}c@{}} RMS motion of payload  \\ during float in tilt ($ ^{\circ}/s$) \end{tabular} &  0.01 \\
\hline
\rule[-1ex]{0pt}{3.5ex} \begin{tabular}{@{}c@{}} Average temperature  \\ inside payload\end{tabular} & 34.35$^{\circ}$C  \\
\hline
\rule[-1ex]{0pt}{3.5ex} \begin{tabular}{@{}c@{}} Average acceleration \\ of payload (X axis)\end{tabular} &  0.176 g \\
\hline
\rule[-1ex]{0pt}{3.5ex} \begin{tabular}{@{}c@{}} Average acceleration \\ of payload (Y axis)\end{tabular} &   0.0036 g\\
\hline\rule[-1ex]{0pt}{3.5ex} \begin{tabular}{@{}c@{}} Average acceleration \\ of payload (Z axis)\end{tabular} &   1.026 g\\
\hline
\end{tabular}
\end{center}
\end{table}

\begin{enumerate}
\item During the flight the payload reached the maximum height of 31.2 km, where the outside temperature 
was $-32^{\circ} $C. The temperature inside the payload stayed above $0^{\circ} $C. This shows that the 
electronic components inside the payload were thermally insulated.
\item The stratospheric conditions during the TIFR flight at float are more stable than the near-surface 
conditions we have experienced during our previous tethered launches at the IIA, and comparable to our 
previous stratospheric flights. The full analysis is presented in the forthcoming paper (Nirmal et al. 2016).
\end{enumerate}

\section{Acknowledgements}

Part of this research has been supported by the Department of Science and Technology (Government of India) under Grant IR/S2/PU-006/2012. We thank Prof.~D.~K.~Ojha for allowing us to use our IMU in their flight. We also thank the staff of the TIFR 
Balloon Facility, Hyderabad, for sharing the GPS and flight information. 

\section{References}
\noindent
Fesen, R. \& Brown, Y. {\it A method for establishing a long duration, stratospheric platform for astronomical research}. 2015,  Experimental Astronomy, 39, 475 \\
\\
\noindent
Gruner.~T.~D., Olney, D.~J. \& Russo, A.~M. {\it Measurements of Load Train Motion on a Stratospheric Balloon Flight}. 2005, NASA Goddard Space Flight Center  Technical Reports Server (NTRS) \\
\\
\noindent
Manchanda, R.K., Subba Rao, J.V., Sreenivasan, S. \& Suneelkumar, B. {\it Study of seasonal variation of winds in upper stratosphere over Hyderabad}. 2011, Advances in Space Research, 47, 480–487\\
\\
\noindent
Nayak A., Sreejith A.~G., Safonova M. \& Murthy J. {\it High-Altitude Ballooning Program at the Indian Institute of Astrophysics}. 2013,  Current Science, 104: 708-713\\
\\
\noindent
Nirmal, K., Sreejith, A. G., Mathew, J., Sarpotdar, M., Suresh  Ambily, M. Safonova \& J. Murthy. {\it Pointing System for the Balloon-Borne Telescope}. 2016, Journal of Astronomical Telescopes, Instruments and Systems, submitted \\
\\
\noindent
Robbins, E. \& Martone, M. {\it Recovery System Termination Load Reduction through the use of a Central Load Core}. 1991, AlAA International Balloon Technology Conference, October 8-10, 1991,  Albuquerque, NM, USA\\
\\
\noindent
Safonova, M., Nayak, A., Sreejith, A. G., Joice Mathew, Mayuresh Sarpotdar, S. Ambily, K. Nirmal, Sameer Talnikar, Shripathy Hadigal, Ajin Prakash \& Jayant Murthy. {\it An Overview of High-Altitude Balloon Experiments at the Indian
Institute of Astrophysics}. 2016, Astron. \& Astroph. Trans., Vol. 29, No. 3, in press \\
\\
\noindent
Sreejith, A. G., Mathew, J., Sarpotdar, M., Nirmal, K., Suresh, A., Prakash, A., Safonova, M. \& Murthy, J. {\it Measurement of limb radiance and Trace Gases in UV over Tropical region by Balloon-Borne Instruments -- Flight Validation and Initial Results}. 2016a, Atmos. Meas. Tech. Discuss., in review \\
\\
\noindent
Sreejith, A. G., Mathew, J., Sarpotdar, M., Nirmal, K., Suresh, A., Prakash, A., Safonova, M. \& Murthy, J. {\it Balloon UV Experiments for Astronomical and Atmospheric Observation}. 2016b, Proc. of SPIE, SPIE Astronomical Telescopes + Instrumentation 2016, Edinburgh, UK

\end{document}